\documentclass[aps,superscriptaddress,showpacs,
 reprint]{revtex4-1} 
 \usepackage[utf8]{inputenc}
 \usepackage{amsmath,amsfonts,amssymb}
 \usepackage{pifont}
 \usepackage{rotating}
 \usepackage{graphicx}    
 \usepackage{epstopdf}
 \usepackage{color}
 \usepackage{cancel}
 \usepackage[caption=false]{subfig}
 \usepackage{soul}
 
 \usepackage{hyperref}
 \definecolor{dark-red}{rgb}{0.9,0.15,0.15}
 \definecolor{dark-blue}{rgb}{0.15,0.15,0.4}
 \definecolor{medium-blue}{rgb}{0,0,0.5}
 \hypersetup{
 	colorlinks, linkcolor={black},
 	citecolor={dark-blue}, urlcolor={medium-blue}
 }

\begin{document} 
\title{Susceptibility anisotropy and absence of ferroelectric order in the Kitaev spin liquid candidate Na$_{2}$Co$_{2}$TeO$_{6}$}

\author{C. Dhanasekhar}
\email{dsekhar21iitb@gmail.com}
\affiliation{Department of Physics, Indian Institute of Technology Bombay, Mumbai 400076, India}
\affiliation{Department of Physics, National Sun Yat-sen University, Kaohsiung 80424, Taiwan}
\affiliation{Center of Crystal Research, National Sun Yat-sen University, Kaohsiung 80424, Taiwan}
\author{Monika Jawale}
\affiliation{Department of Physics, Indian Institute of Technology Bombay, Mumbai 400076, India}
\author{Rahul Kumar}
\affiliation{School of Advanced Materials, Chemistry and Physics of Materials Unit,
Jawaharlal Nehru Centre for Advanced Scientific Research, Bangalore 560064, India}
\author{D. Chandrasekhar Kakarla}
\affiliation{Department of Physics, National Sun Yat-sen University, Kaohsiung 80424, Taiwan}
\author{Sagar Mahapatra}
\affiliation{Department of Physics, Indian Institute of Science Education and Research, Pune 411008, India}
\author{ Mitch M. C. Chou}
\affiliation{Center of Crystal Research, National Sun Yat-sen University, Kaohsiung 80424, Taiwan}
\author{A. Sundaresan}
\email{sundaresan$@$jncasr.ac.in}
\affiliation{School of Advanced Materials, Chemistry and Physics of Materials Unit,
Jawaharlal Nehru Centre for Advanced Scientific Research, Bangalore 560064, India}
\author{H. D. Yang}
\email{yang@mail.nsysu.edu.tw}
\affiliation{Department of Physics, National Sun Yat-sen University, Kaohsiung 80424, Taiwan}
\affiliation{Center of Crystal Research, National Sun Yat-sen University, Kaohsiung 80424, Taiwan}
\author{A.V.Mahajan} 
\email{mahajan$@$phy.iitb.ac.in}
\affiliation{Department of Physics, Indian Institute of Technology Bombay, Mumbai 400076, India}
 
 
\begin{abstract}

We report the magnetic, magnetodielectric, and electric polarization properties of single crystals of the Co-based Kitaev Spin Liquid (KSL) candidate Na$_{2}$Co$_{2}$TeO$_{6}$ (NCTO). The sample shows magnetic transitions at 26\,K, 16\,K, and 5\,K, consistent with the literature. The magnetic measurements along and perpendicular to the Co-honeycomb planes show a strong anisotropy in susceptibility and in Curie-Weiss (C\,-\,W) temperatures. The experimental anisotropic C-W temperatures of NCTO qualitatively match with the theoretical C-W temperatures, calculated by using the HK$\Gamma\Gamma^{'}$ model [C. Kim et al. Journal of Physics: Condensed Matter \textbf{34}, 045802 (2021)]. We find from our temperature and field dependent  dielectric and pyroelectric ($I_{p}$) current studies ($H \vert \vert \,ab$ and  $E \bot\,ab$) that our single crystal  NCTO samples do not have a  finite electric polarization below 100\,K. These $I_{p}$ studies confirm the absence of a magneto-electric coupling and electric polarization properties in the title compound and suggest that the  zig-zag AFM structure is more favorable than the triple-Q structure with AFM Kitaev interactions. 

\end{abstract}
		
\date{\today}

\maketitle

\section{Introduction}

Similar to the geometrically frustrated magnetic materials, the competition among spin-spin interactions can also induce novel magnetic ground states. In this context, a representative example is the Kitaev spin liquid (KSL), which is a type of quantum spin liquid realized on a honeycomb lattice, where the anisotropic interactions driven by spin-orbit coupling play a major role \cite{Takagi2019,Trebst,winter}. In the KSL state the spins are fractionalized into Majorana fermions, which could serve as a basis for quantum information and quantum computing applications \cite{OBrien,Motome,Hermanns}. Motivated by the theoretical work of Jackeli and Khaliullin \cite{Jackeli} materials-based research has been extensively carried out on low-spin Ru$^{3+}$ ($4d^{5}$) and Ir$^{4+}$ ($5d^{5}$) compounds, such as quasi-two-dimensional (2D) honeycomb magnets A$_{2}$IrO$_{3}$ (A = Li, Na), Ag$_{3}$LiIr$_{2}$O$_{6}$, H$_{3}$LiIr$_{2}$O$_{6}$, and $\alpha$-RuCl$_{3}$, three dimensional (3D) hyperhoneycomb magnet $\beta$-Li$_{2}$IrO$_{3}$, and 3D stripy honeycomb magnet $\gamma$-Li$_{2}$IrO$_{3}$ \cite{Jackeli,Singh,Atasi,Bahrami,winter,bachhar}. The honeycomb layered cobaltates (A$_{3}$Co$_{2}$SbO$_{6}$ and A$_{2}$Co$_{2}$TeO$_{6}$ A= Na, Li, Ag) were recently predicted to be valid candidates for hosting the KSL state. Although the spin-orbit coupling is small in 3$d$-Co as compared to Ir and Ru, the extra spin in $e_{g}$ orbitals of 3$d$-Co materials plays a role in the canceling out of the Heisenberg exchange process and enhancing the Kitaev spin interactions \cite{Khaliullin,Songvilay,Sanders,Stephen}.

The Kitaev model is based on bond-dependent anisotropic interactions between spins on a honeycomb lattice, leading to highly anisotropic magnetic and thermal properties \cite{Yan,Gillig,Lampen}. Materials that are candidates for the KSL state often have competing magnetic phases due to the presence of additional interactions beyond the pure Kitaev limit. Thus, studying and understanding the anisotropic magnetic properties helps to distinguish the KSL state from other magnetic phases. A strong anisotropy in the magnetic susceptibilities measured along the parallel ($H \vert \vert \,ab$) and perpendicular ($H \bot\,ab$) to the honeycomb plane have been reported in 4$d$ - and 5$d$-KSL materials \cite{Kubota,Sears,Baenitz,Yogesh}. Further, various theoretical models have been built to understand the experimental anisotropic magnetic susceptibilities in 4$d$ and 5$d$, KSL materials and these studies highlight the importance or contributions  of the other exchange interactions in these materials \cite{Andrade,Janssen2017,Singh2017,Winter2016}.

In KSL materials rigorous studies have been performed to understand the effects arising from the spin degrees of freedom, however the charge degrees of freedom  are often ignored. Combining the spin and charge degrees of freedom (such as electric polarization) in the KSL materials might open up the field to new  applications \cite{Pereira,Chari,Noh,Geirhos,zheng2024}. To understand such correlations, dielectric  studies on single crystals of $\alpha$-RuCl$_{3}$ have been performed, which show a dielectric anomaly close to the magnetic ordering temperature. However, detailed dieletric studies on various single crystals of $\alpha$-RuCl$_{3}$ suggest that  the observed dielectric anomaly likely originates from  the interfaces between stacking layers \cite{Mi}. The pyroelectric current studies on the $\alpha$-RuCl$_{3}$ single crystals  show the appearance of  ferroelectric polarization in the Kitaev paramagnetic state. The electric polarization further  persists in the long-range AFM  state and also survives in the field-induced magnetically disordered  state. The observed electric polarization in $\alpha$-RuCl$_{3}$ is explained by considering the charge fluctuations and distortions in Ru honeycomb plane \cite{Xinrun}.

In this respect, the recent exploration of  cobalt-based honeycomb KSL candidate Na$_{2}$Co$_{2}$TeO$_{6}$ is  of special interest, as the magnetic ground state (zig-zag vs. triple-Q) \cite{Chen,Yao,Gaoting,Bera,Wilhelm,Lee} and reported electric polarization properties are not consistent \cite{Chaudhary,Mukherjee,zhang}. Initially, Chaudhary $\it{et \, al.}$  (\cite{Chaudhary}) and  Mukherjee $\it{et \, al.}$ \cite{Mukherjee} reported the electric polarization properties  in polycrystalline Na$_{2}$Co$_{2}$TeO$_{6}$ (NCTO) samples below 100\,K. Although the magnetic properties of these two studies match well, the ferroelectric ground state is clearly different in the two cases. Studies in Ref. \cite{Chaudhary} show the absence of coupling between the spin and charge degrees of freedom, whereas in Ref. \cite{Mukherjee}, they have claimed the presence of  a finite coupling. More recently, Zhang et al. \cite{zhang} reported the absence of electric polarization along the $a^{\ast}$-axis in NCTO single crystals by applying the magnetic field ($H$) along the $a$-axis.  Based on the absence of electric polarization, Zhang et al. \cite{zhang} suggested that the zig-zag spin structure is more likely in NCTO in zero magnetic field than the triple-Q magnetic structure. The KSL materials show highly anisotropic physical properties. Further, as the crystal structure of these materials consists of layers, they are prone to stacking faults and defects which can cause differences in the physical properties, as reported in the well studied KSL candidate  $\alpha$-RuCl$_{3}$. Thus, it is essential to revisit magnetic, dielectric, and ferroelectric  properties of NCTO, so as to clarify the  ground state of the title compound.

\section {Experimental Details}

The plate-like Na$_{2}$Co$_{2}$TeO$_{6}$ single crystals were grown by the flux method similar to that in literature \cite{Yao,Yao2020}. The orientation of the NCTO crystals were measured using an X-ray Laue diffractometer (Photon Science, UK) with a tungsten target. To understand the  Co$^{2+}$ valence state in the prepared NCTO samples, the X-ray absorption spectra (XAS) studies were performed in a total electron yield (TEY) mode at the BL20 beamline of the National Synchrotron Radiation Research Centre (NSRRC), Taiwan.  The temperature dependent magnetic measurements were carried out using a Magnetic Property Measurement System, Quantum Design, USA. Temperature and magnetic field dependent dielectric measurements were performed  using the LCR meter (Agilent E4980A). The pyroelectric current ($I_{p}$) was  measured using Keithley 6517A electrometer under conventional electric poling and bias poling electric field methods during in warming temperature cycles \cite{Sundaresan,Ajay,CDS}. All the dielectric and pyroelectric current measurements were carried out with the help of  homemade multifunctional probe, attached with the PPMS. For the single crystals the area of the electrodes for the honeycomb plane was 2 mm$^{2}$ and the thickness of the  single crystal sample was about 0.21 mm. For the dielectric and pyroelectric current measurements on the  single crystals, the $E \bot\,ab$ and $H \vert \vert \,ab$ directions were used.

\section {Result and discussion}

\subsection{Room temperature X-ray diffraction and X-Ray Absorption (XAS) studies}

\begin{figure}[hbt!]
\includegraphics[scale=0.6]{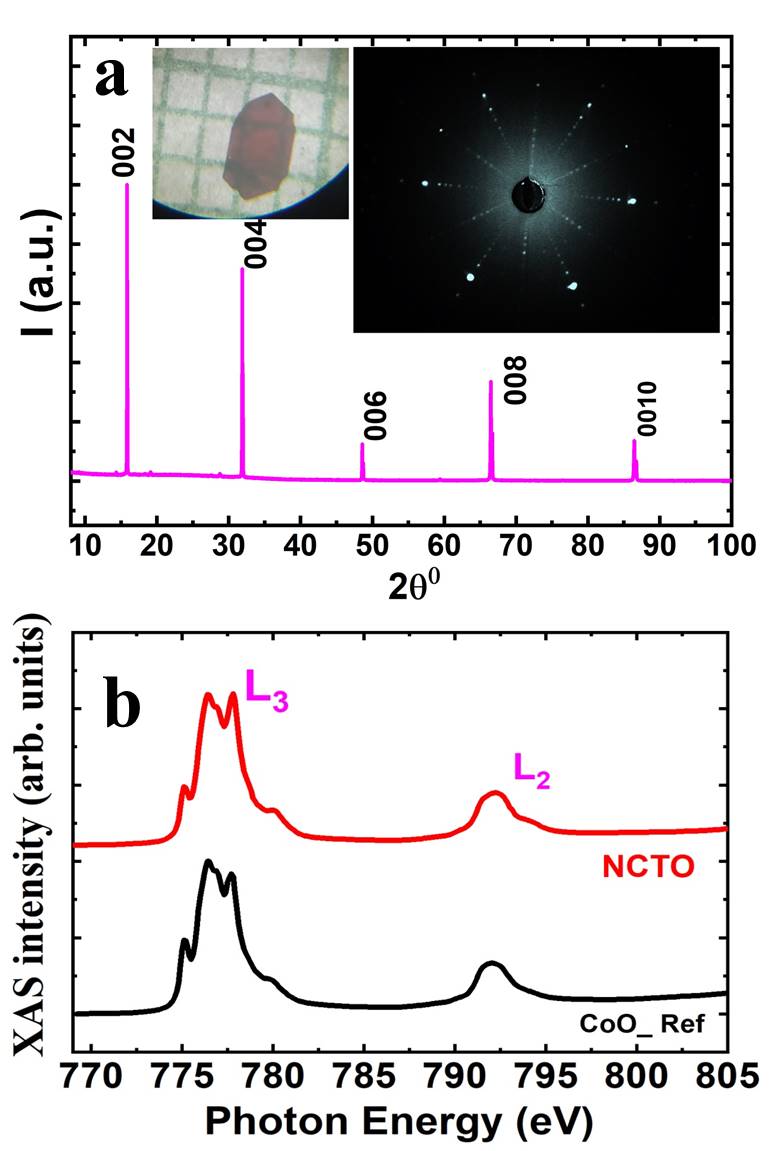}
\centering
\caption {(a) Room temperature XRD pattern of a large hexagonal NCTO crystal is shown; insets show the Laue back diffraction pattern and a picture of the NCTO crystal. (b) Co 2p XAS spectra of the NCTO samples along with CoO reference spectra are shown.}
\label{XRD_NCTO}	
\end{figure}

The as-grown crystals of NCTO are thin plates with a hexagonal shape and are several millimeter in size, as shown in the inset of Fig.\,\ref{XRD_NCTO}(a).The thin plates are normal to the crystallographic $c$-axis, as confirmed by Laue back-diffraction pattern [inset of Fig.\,\ref{XRD_NCTO}(a)]. Further, the room temperature XRD pattern from the large hexagonal plate like surfaces of the NCTO single crystals is shown in the main panel of the Fig.\,\ref{XRD_NCTO}(a), which only shows the (00l) Bragg peaks. Fig.\,\ref{XRD_NCTO}(b) shows the Co 2$p$ XAS spectra of NCTO single crystals along with the comparison of the CoO. The Co 2p XAS spectrum shows two features at L$_{3}$ $(h\nu = 775\,–\,785\,eV)$ and L$_{2}$ $(h\nu = 790\,–\,800\,eV)$ edges. The Co L$_{3}$ edge consists of different sharp peak structures and also has shoulder on the higher photon energy side. On the other hand, the L$_{2}$ edge has a broad structure centered at 792 eV. Because the XAS spectral shapes strongly depend on the valence state, the valency of the Co ions can be determined by comparing with the reference XAS spectra. The Co 2$p$ XAS spectra of NCTO samples are quite similar to the spectrum of CoO, suggesting that the Co ions in NCTO and CoO have an identical valence state, i.e., the divalent Co ions in CoO$_{6}$ octahedra have the high-spin configuration t$_{2g}$$^{5}$e$_{g}$$^{2}$ \cite{Burnus}.

\subsection{Magnetization studies}

The temperature ($T$) dependence of the magnetic susceptibility measured in a 100 Oe field for a NCTO single crystal is shown in  Fig.\,\ref{NCTO_MT_final}(a). In these measurements,  $\chi_{c}$ denotes the susceptibility measured with the  applied field perpendicular to the honeycomb ($ab$) plane ($H \bot\,ab $) whereas   $\chi_{ab}$ denotes the susceptibility measured parallel to the honeycomb  plane ($H \vert \vert \,ab$). The $\chi_{ab}$ and  $\chi_{c}$  show  different features below 30\,K.  Close to 26\,K,  $\chi_{ab}$ shows a sharp increase whereas $\chi_{c}$ shows a decrease. Upon further decreasing the temperature,  $\chi_{ab}$ shows a broad maximum centered around 20\,K,   and an upturn below 5\,K. In contrast to $\chi_{ab}$, the $\chi_{c}$ shows a continues decrease below 26\,K. To visualize  the various magnetic features clearly, we have plotted the first derivative of the susceptibilities and is shown in the inset of the  Fig.\,\ref{NCTO_MT_final}(a). Based on the previous studies, the transition at 26\,K  is assigned to  long-range antiferromagnetic order ($T_{N1}$) and the transitions at 16\,K and at 5\,K are assigned to the onset of spin canting/spin reorientation \cite{Gaoting}. The difference in $\chi_{ab}$ and  $\chi_{c}$ above and below  26\,K suggests that the magnetic susceptibility is highly anisotropic in nature. For $T> T_{N}$,  $\chi_{ab}$ is slightly greater than $\chi_{c}$ and this difference gets smaller on decreasing $T$, and for $T< T_{N}$, $\chi_{ab}$ is less than $\chi_{c}$. Fig.\,\ref{NCTO_MT_final}(b) shows the  $M(H)$ curves of NCTO single crystals measured along the   $H \bot\,ab $ ($\chi_{c}$) and  $H \vert \vert \,ab$ ($\chi_{ab}$ )  directions, respectively. At 2\,K,  $M_{ab}$ shows a hysteresis with the magnetic field between 30\,-\,70\,kOe  and also shows an upturn with increasing field. Such an upturn in $M(H)$ suggests a field-induced magnetic phase transition. At 2\,K, under maximum field of 7\,T the obtained magnetization is 1.25 $\mu_{B}$ per {Co$^{2+}$ along the $\chi_{ab}$ and 0.65 $\mu_{B}$ per {Co$^{2+}$ along the $\chi_{c}$. The  $\chi \,vs.T$ and $M(H)$ behavior of the NCTO matches with the literature \cite{Bera,Lefran,Xiao}.

\begin{figure}[hbt!]
\includegraphics[scale=0.55]{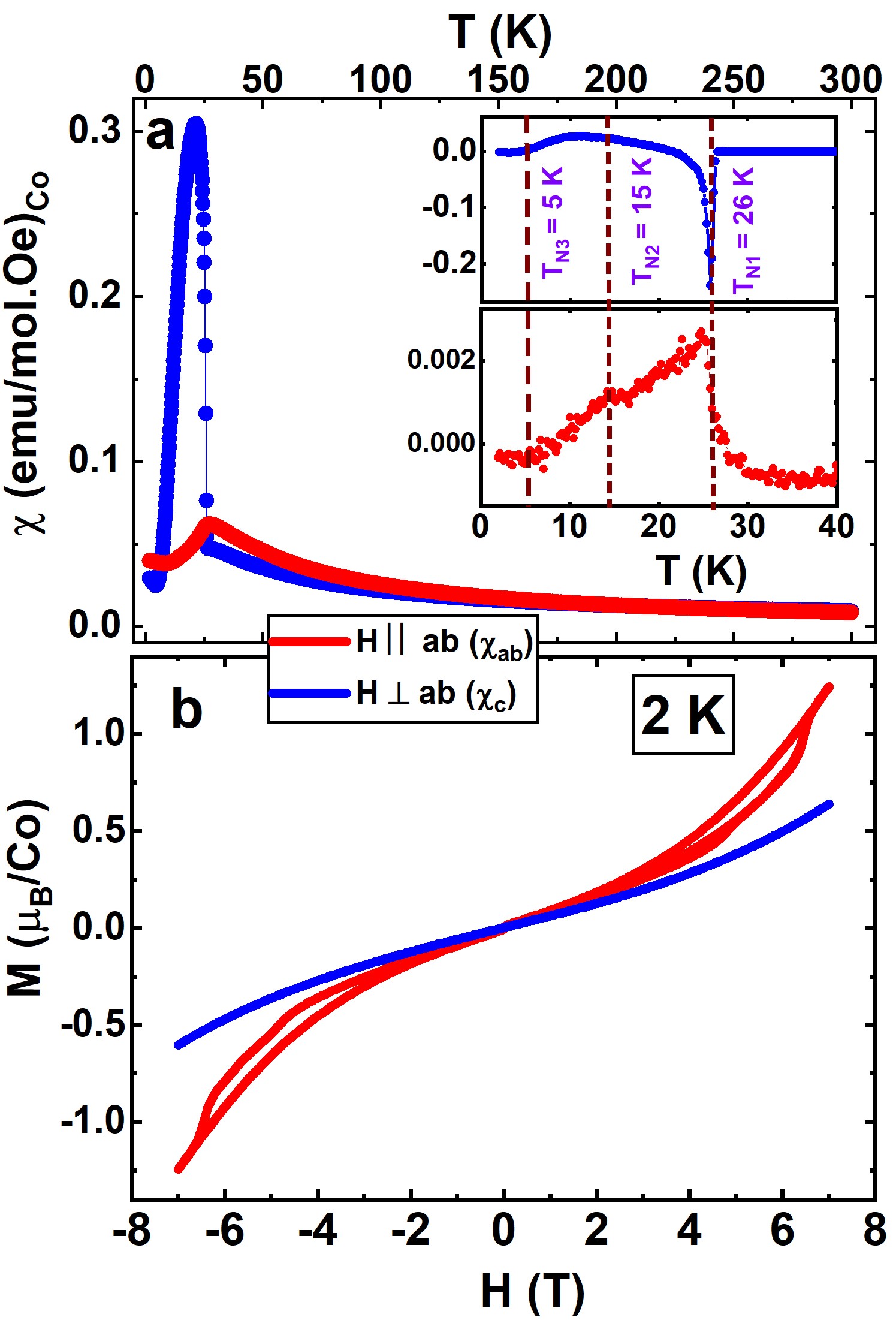}
\centering
\caption{The main figure in (a) shows $\chi$ vs $T$ for the NCTO single crystal sample measured in a 100 Oe field directed  perpendicular and parallel to the honeycomb (ab) plane.  The inset shows the various magnetic transitions obtained from the first derivative of $\chi$ with $T$.  The (b) $M$ vs. $H$ curves of NCTO single crystals measured at 2\, K }
\label{NCTO_MT_final}
\end{figure}

\maketitle \begin{table*}[th!]
\centering
\begin{tabular}{c c c c c c c c c} 
\hline
$T$-range \,&\, $\chi_{0}$ ($\times 10^{-4} emu/mol.Oe. _{Co}$) \,&\, $\theta_{CW}$ (in K) \,&\, $\mu_{eff}$ $(\frac{\mu_{B}}{Co^{2+}})$ \,&\, Ref \ \\ [0.5ex] 
\hline\hline
200\,K-300\,K\,&\, $-$ \,&\, $\theta a_{^{\ast}} $=$\,12.5(6)$,$\theta_{c}$=$ $-$93.8(10)$ \,&\,  $\mu_{a_{^{\ast}}}$  $=$ 5.69(8), $\mu_{c}$ $=$ 5.31(6) \,&\,\cite{Yao2020} \,\\ 
100\,K-300\,K \,&\, $-$ \,&\, $\theta_{ab} $=$\,$-$9$, $\theta_{c}$=$ $-$139$  \,&\,$\mu_{ab}$  $=$ 5.99, $\mu_{c}$ $=$ 5.98 \,&\,\cite{Xiao} \,\\ 
100\,K-300\,K \,&\,$\chi_{0\,c}$ $=$\, $+$15.7(8)\, \,&\, $\theta_{c}$=$\,  $-$17.62(35)$  \,&\ $\mu_{c}$ $=$ 4.33(25) \,&\,this study \,\\ 
\,&\,\,$\chi_{0\,ab}$ $=$\, $-$9.7(9)  \,&\, $\theta_{ab} $=$\,$-$4.77(29)$ \,&\,$\mu_{ab}$  $=$ 4.77(25) \,&\, \,\\ 
150\,K-300\,K \,&\,$\chi_{0\,c}$ $=$\, $-$5.2 \, \,&\, $\theta_{c}$=$\,  $-$113$  \,&\ $\mu_{c}$ $=$ 5.13 \,&\,\cite{zhang2024}\,\\ 
  &\,\,$\chi_{0\,a}$ $=$\, $-$20.0  \,&\, $\theta_{a} $=$\,$+$2.11$ \,&\,$\mu_{a}$  $=$ 5.7 \,&\, \,\\ [1ex]
110\,K-300\,K \,&\, $-$ \, \,&\, $\theta a_{^{\ast}} $=$\,13.2(2)$,$\theta_{c}$=$ $-$104.5(2)$\,&\ $\mu_{a_{^{\ast}}}$  $=$ 5.35, $\mu_{c}$ $=$ 5.37 \,&\,\cite{Lee}\,\\ [1ex] 
\hline
\end{tabular}
\caption{Comparison of the Curie-Weiss fitting results of  NCTO single crystals.}
\label{table:1}
\end{table*}

The magnetic susceptibility is further analysed using the Curie-Weiss (C\,-\,W) law $\chi$ = $\chi_{0}+C/(T-\theta_{CW})$ where $C$ is the Curie constant, $\chi_{0}$ is the temperature independent susceptibility, and $\theta_{CW}$ is the Curie-Weiss temperature. The C\,-\,W fitting for $\chi_{ab}$ and  $\chi_{c}$ in the range of the 100\,K - 300\,K gives the temperature independent susceptibilities of $-$9.9(0.2)$\times10^{-4}$\,emu/mol.Oe ($\chi_{0, ab}$), and $+$15.7(0.1)$\times10^{-4}$\,emu/mol.Oe ($\chi_{0,c}$), respectively. The $\chi_{0}$ is the sum of the core diamagnetic susceptibility ($\chi_{dia}$ ) and Van Vleck susceptibility ($\chi_{VV}$). The sum of the core diamagnetic susceptibility ($\chi_{dia}$) of the individual ions give a total $\chi_{dia}$ for NCTO to be $-$0.63$\times10^{-4}$\,emu/mol.Oe. Subtracting  the  $\chi_{dia}$ from $\chi_{0}$ gives us the $\chi_{VV}$ values of $+$16.3$\times10^{-4}$\,emu/mol.Oe. and $-$9.3$\times10^{-4}$\,emu/mol.Oe along the $\chi_{c}$ and $\chi_{ab}$ directions. The $\chi_{VV}$ of the Co$^{2+}$ is reported as $+$8.27$\times10^{-3}$\,emu/mol \cite{Tanaka}. The obtained $\chi_{VV}$ from an analysis of $\chi_{c}$ is close to the expected value from literature whereas the $\chi_{VV}$ from $\chi_{ab}$ analysis gives negative values. This is likely due to an incorrect subtraction of the diamagnetic contribution from the sample holder for this orientation. Here, we note,that in literature the $\chi_{VV}$ of Co$^{2+}$ is calculated from $M(H)$ curves at high magnetic fields. After subtracting the $\chi_{0}$ from $\chi$, we find that 1/($\chi-\chi_{0}$) varies linearly with temperature down to 70\,K, which is shown in the Fig.\,\ref*{MH_NCTOf}.

\begin{figure}[hbt!]
\includegraphics[scale=0.5]{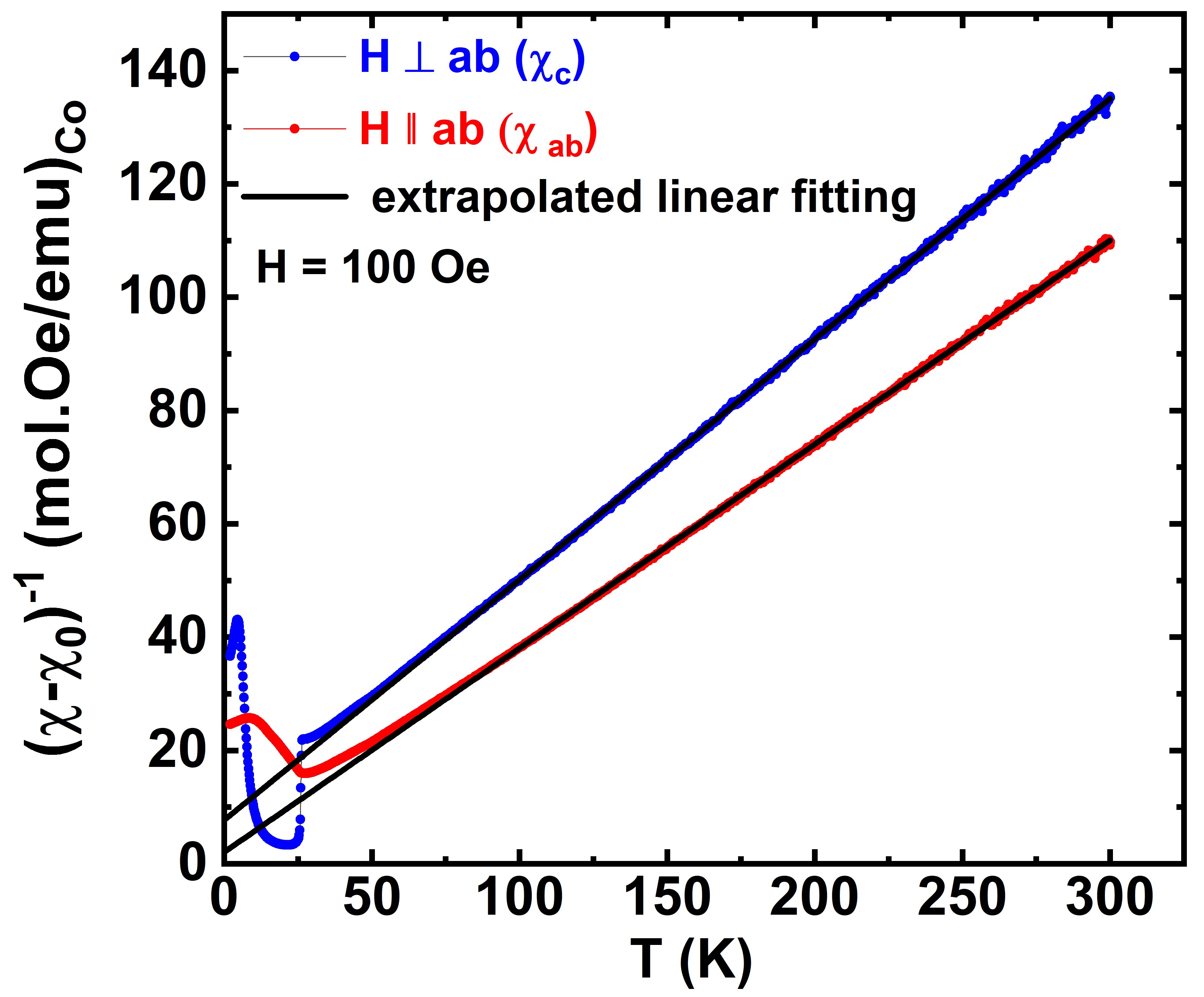}
\centering
\caption{Display of 1/($\chi-\chi_{0}$) vs. $T$ for single crystal  NCTO along with C-W fitting. The C-W fitting is performed in the  $T$ range of 100\,K\,-\,300\,K.}
\label{MH_NCTOf}
\end{figure}

The effective paramagnetic moments obtained from the C-W fit of the single crystal are $\mu_{eff} = 4.33(25) \,\frac{\mu_{B}}{Co^{2+}}$ and $\mu_{eff} = 4.77(25) \,\frac{\mu_{B}}{Co^{2+}}$ for $\chi_{c}$ and $\chi_{ab}$, respectively. The C\,-\,W temperatures are $\theta_{c}$ $=$ $-$17.62(35)\,K and $\theta_{ab}$ $=$ $-$4.77(29)\,K respectively. From the $\mu_{eff}$  ($=$ $g$ $\sqrt{j(j+1)}$, $j$ = 1/2) the calculated $g$ factor anisotropies for the NCTO single crystals are $g_{c}$= 5 and $g_{ab}$=5.46. The $g$-factors obtained here are higher than the values obtained from the paramagnetic state of HF-ESR and HF-M(H) curves in NCTO samples ($g_{c}$= 2.3 and $g_{ab}$=4.13) \cite{GaotingLin,zhang2024}. However, the $g$-factor anisotropies obtained by us are similar to those obtained from high temperature $\mu_{eff}$ values \cite{Li}. The discrepancy maybe due to a mixing of the   $j = 1/2$ and  $j = 3/2$ states in the 100-300 K temperature range. The observed  experimental $\mu_{eff}$ values are  larger than the spin-only value, which probably indicates a significant contribution from the orbital moment \cite{Viciu,Bera,Lefran,Xiao}.  The  $\mu_{eff}$ values  are consistent with earlier reports, but the $\theta_{CW}$ values differ significantly between various studies, as shown in the Table.I.
 
\subsection{Dielectric and pyroelectric current  studies}

\begin{figure}[hbt!]
\includegraphics[scale=0.6]{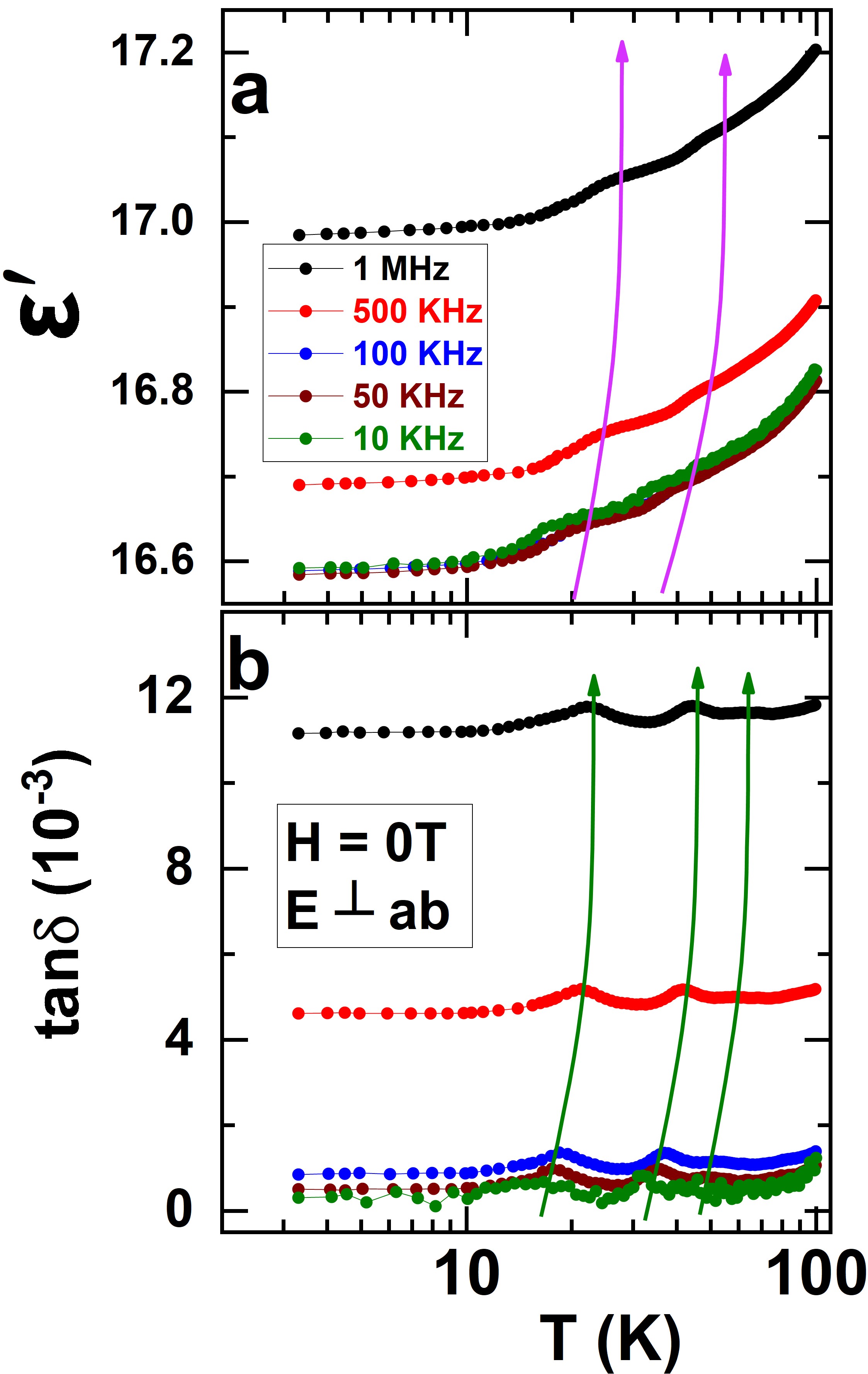}
\centering 
\caption{The (a) $\&$ (b) shows  the logarithmic temperature dependence of the $\epsilon^{'}$ and tan\,$\delta$ of NCTO single crystal measured at different frequencies under $H$=\,0\,T.The electric field is applied perpendicular to the $ab$-plane. The vertical arrows show the frequency and temperature variation in the humps  of the $\epsilon^{'}$ and  peak positions in tan\,$\delta$.}
\label{Dieleltric_SCf}
\end{figure}

Temperature and frequency variation of the dielectric constant  $\epsilon^{'}$ and the loss parameter tan\,$\delta$ of NCTO single crystals measured under $H=0$ and electric field($E_{ac}$ = 1 V) perpendicular to the $ab$ plane ($E \bot\,ab $) is shown in Fig.\,\ref{Dieleltric_SCf}(a $ \& $ b). The $\epsilon^{'}$ and tan\,$\delta$ both show the two frequency dependent broad humps at the $T$ range between 20\,K - 30\,K and 30\,K - 40\,K respectively. Further, tan\,$\delta$ also shows an additional broad hump at the $T$ range between 50\,K - 60\,K. In earlier studies, such frequency dependent features in the NCTO were suggested to arise from freezing of Na$^{+}$ positions as temperature decreases \cite{zhang}. Further, the isothermal dielectric constant, i.e., $\epsilon^{'}(H)$  is also measured at 3\,K, 15\,K and 20\,K by applying the $H \vert \vert \,ab$ , $E \bot\,ab $, and the obtained results are shown in the Fig.\,\ref{Dielectric_MD}. The $\epsilon^{'}(H)$ measured at 3\,K shows  a clear up-turn above 50 kOe and such behavior ties in with the in-plane magnetization measurements where we have noticed spin-flop features above 50 kOe (Fig.\,\ref{NCTO_MT_final}(c)).  Further, the $\epsilon^{'}(H)$ measured at 15\, K and at 20\, K shows additional features, i.e.  $\epsilon^{'}(H)$ shows multiple humps  at different magnetic fields and are shown in Fig.\,\ref{Dielectric_MD}(b and c) with arrow marks. The  $\epsilon^{'}(H)$ also shows hysteresis behavior under increasing and decreasing $H$ and also the hump positions are very sensitive to the $H$ directions.

\begin{figure}
\includegraphics[scale=0.5]{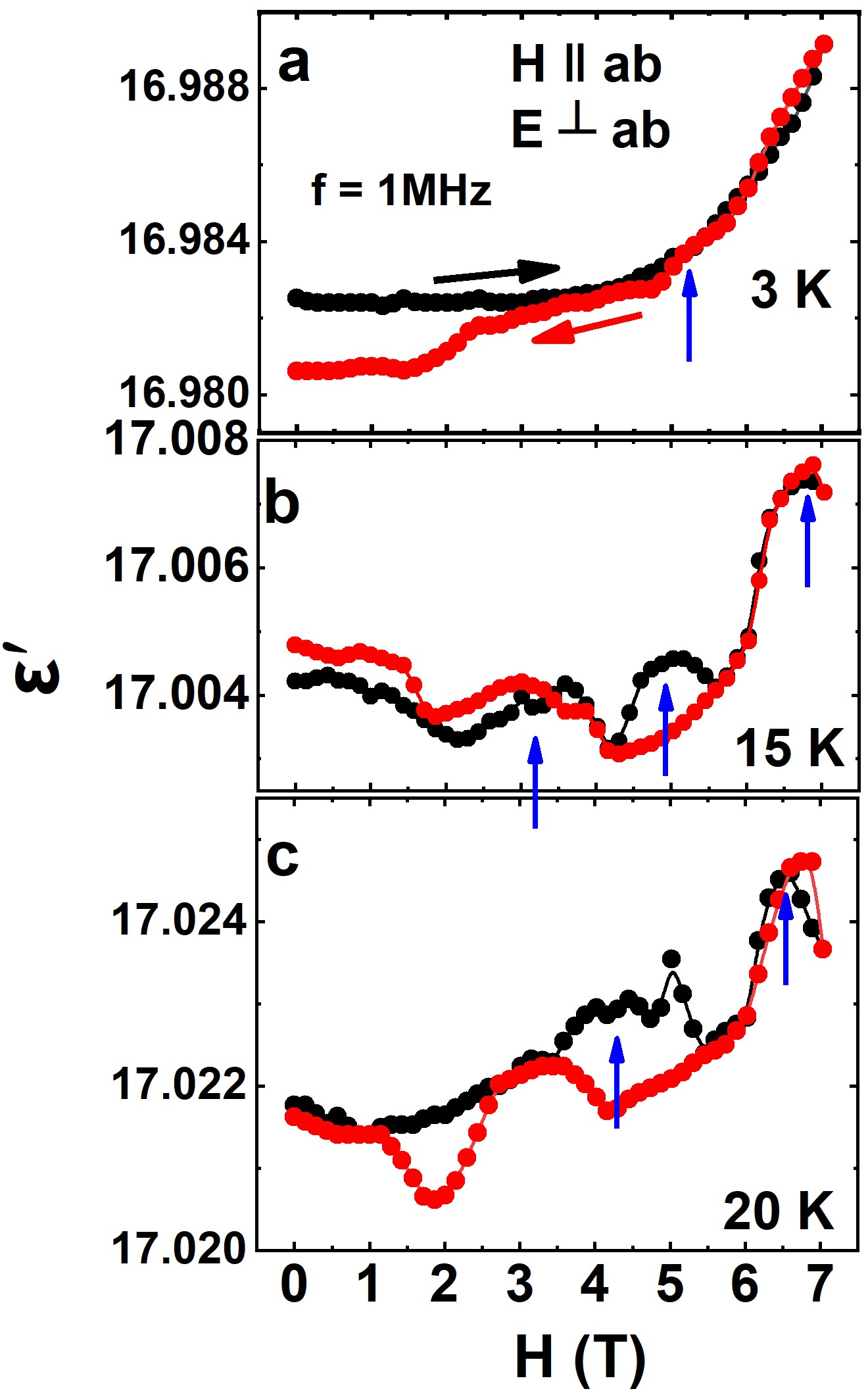}
\centering 
\caption{The figure shows   $\epsilon^{'}(H)$  of NCTO single crystals at different temperatures measured for $H \vert \vert \,ab$ and $E \bot\,ab $ directions.}
\label{Dielectric_MD}
\end{figure}

To further understand if the humps in the temperature dependence of the dielectric constant are associated with any local electric polarization in the NCTO, we have performed the $I_{p}$ ($H \vert \vert \,ab$ and $E \bot\,ab $  conditions) measurements on our single crystals, using electric poling and dc bias poling methods and the obtained results are shown in the Fig.\,\ref{Pyro_Final}. The obtained  $I_{p}$ results show absence of sharp asymmetric $I_{p}$ peaks, which generally appear for the pure spin driven ferroelectric materials. Here, we note that the earlier dielectric and  pyroelectric studies on single crystals of NCTO measured under for directions ($H \vert \vert \,a$ and $E \vert \vert \,a^{\ast}$) also show similar $\epsilon^{'}$ and tan\,$\delta$ features and also the  absence of $I_{p}$ peaks \cite{zhang}. These results confirm the absence of ferroelectricity and  magnetoelectric phenomena in single crystal of NCTO.

\begin{figure}
\includegraphics[scale=0.6]{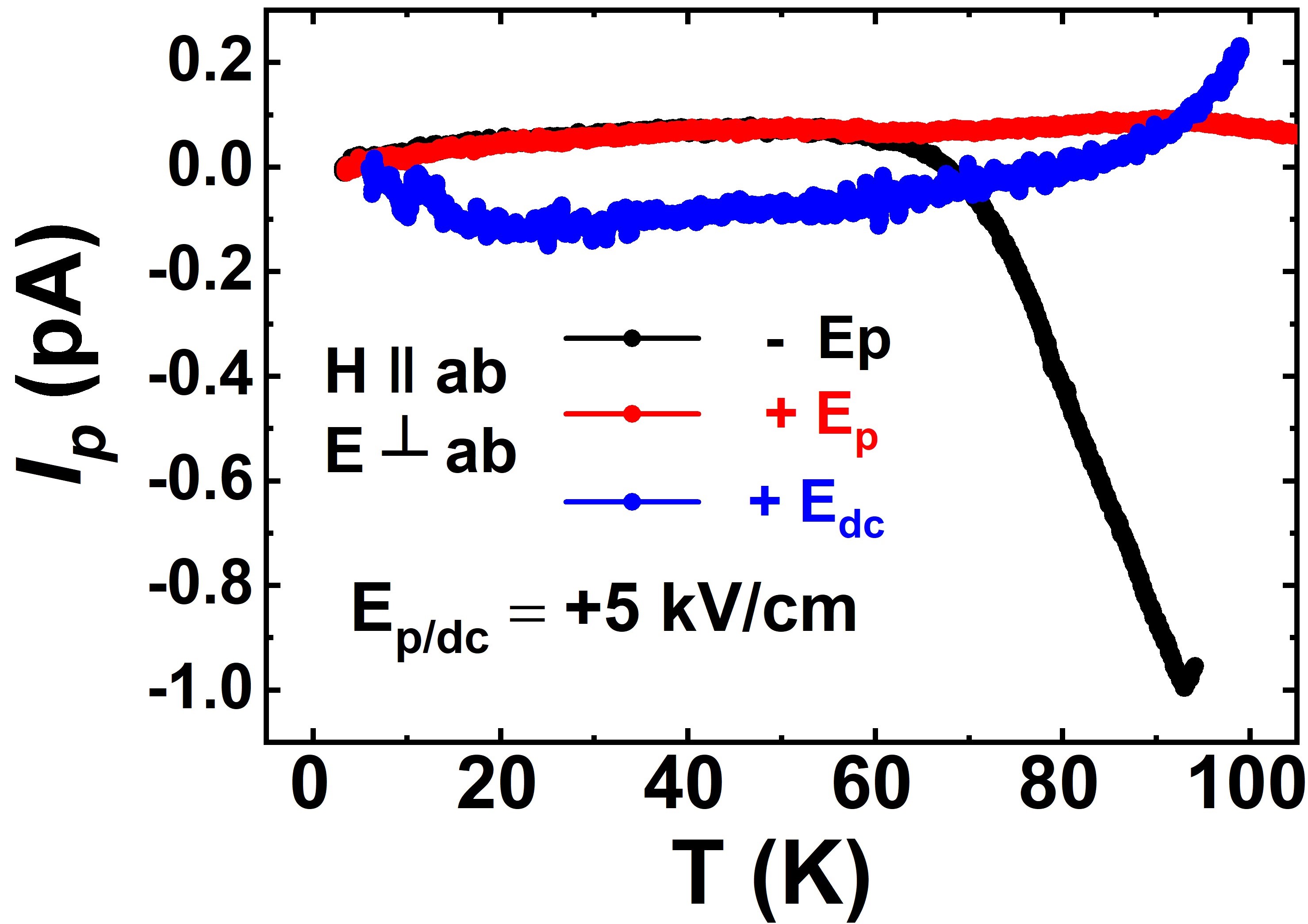}
\centering 
\caption{The $T$  dependence of  $I_{p}$ obtained on single crystal samples under normal poling ($ E_{p}$) and dc bias poling ($ E_{dc}$) conditions measured for $H \vert \vert \,ab$ and $E _{p/dc} \bot\,ab $ directions.} 
\label{Pyro_Final}
\end{figure}

\subsection{Discussion}

\begin{table*}[th!]
\centering
\begin{tabular}{c c c c c c c c} 
\hline
 $J_{1}$ \,&\, $J_{3}$ \,&\, $K$ \,&\, $\Gamma$  \,&\, $\Gamma^{'}$  \,&\, $\theta_{CWab}$ (in K)\,&\,   $\theta_{CWc}$ (in K) \,&\, Ref \ \\ [0.5ex] \hline\hline
 $-$1.5 \,&\, $+$1.5\, \,&\,$+$3.3 \,&\, $-$2.8 \,&\,$+$2.1 \,&\,$-$5.5 \,&\,$-$17.6&\,\cite{Kim_2022}\\ 
 $-$0.1 \,&\, $+$0.9\, \,&\,$-$9 \,&\, $+$1.8 \,&\,$+$0.3 \,&\,$+$26.1 \,&\,$+$5.2 &\,\cite{Songvilay}\\ 
 $-$0.2 \,&\, $+$1.2\, \,&\,$-$7 \,&\, $-$0.02 \,&\,$-$0.23 \,&\,$+$10.3 \,&\,$+$14.1 &\,\cite{Samarakoon}\\ 
 $-$3.2 \,&\, $+$1.2\, \,&\,$+$2.7 \,&\, $-$2.9 \,&\,$+$1.6 \,&\,$+$10.4 \,&\,$+$7.8 &\,\cite{Samarakoon}\\ 
 $-$0.2 \,&\, $+$1.6\, \,&\,$-$7 \,&\, $+$0.5 \,&\,$+$0.15 \,&\,$+$10.4  \,&\,$+$3.4 &\,\cite{Sanders}\\ 
 $-$3.5 \,&\, $+$1.4\, \,&\,$+$3.2 \,&\, $-$3 \,&\,$+$2 \,&\,$+$11.8 \,&\,$+$3.1 &\,\cite{Sanders}\\ 
 $-$2.32 \,&\, $+$2.5\, \,&\,$+$0.125 \,&\, $+$0.125 \,&\, 0 \,&\,$-$1.5  \,&$-$2.6 &\,\cite{GaotingLin}\\[1ex]    
\hline
\end{tabular}

\caption{ The exchange parameters (in meV) of the HK$\Gamma$$\Gamma{'}$ model extracted from fitting of inelastic neutron scattering data. $J_{1}$ and $J_{3}$ stand for the isotropic exchange between first and third nearest neighbors,K is the Kitaev parameter, $\Gamma$  and $\Gamma^{'}$  are  the off-diagonal exchange energy terms. The $\theta_{CWab}$, $\theta_{CWc}$ are calculated using the  equations(1) and (2). The positive sign corresponds to AFM and the negative sign corresponding to the FM. Ref \cite{Samarakoon} includes  the next nearest neighbor ($J_{2}$) and interlayer ($J_{c}$) coupling interactions.}
\label{table:1}
\end{table*}

The anisotropy in Curie-Weiss temperature of KSL materials  has an intriguing consequence in view of the theoretical models. The anisotropic susceptibility in honeycomb Kitaev materials  have been modeled using the  HK$\Gamma\Gamma^{'}$ model and Monte Carlo (MC) simulations and high temperature expansion studies \cite{Khaliullin,Andrade}. According to this model the in-plane ($\theta_{CWab}$) and out of plane ($\theta_{CWc}$) C-W temperatures can expressed as bellow;
\\ 
\newline
$\theta_{CWab}$  = $-$\,c/3 $\times$ [3($J_{1}$+$J_{3}$)+$K$-($\Gamma$+2$\Gamma^{'}$)]\,\,\,\,\, (1)
\\
\newline
$\theta_{CWc}$ = $-$\,c/3 $\times$ [3($J_{1}$+$J_{3}$)+$K$+2($\Gamma$+2$\Gamma^{'}$)]\,\,\,\,\, (2)
\\
\newline
where c $=$ $S(S+1)$ for quantum spins, and for classical spins c $=$ $S^{2}$, $J_{1}$ and $J_{3}$ are the first and third nearest neighbor Heisenberg exchange energies, $K$ is the Kitaev energy term, $\Gamma$ and $\Gamma^{'}$ are the off-diagonal exchange energy terms. Extensive neutron scattering studies have been performed on NCTO  and the obtained $J_{1}$, $J_{3}$ , $K$, $\Gamma$, and $\Gamma^{'}$ energies using the HK$\Gamma\Gamma^{'}$ model are shown in Table II. Using these values we calculated (considering $S=1/2$ and $c$ $=$ $S(S+1)$,\cite{Khaliullin}) the  $\theta_{CW \vert \vert}$ and $\theta_{CW \bot}$ from the equations (1) and (2), and the obtained C-W temperatures are also shown in the Table II. Our C-W temperatures are close to the ones inferred by Kim et al., \cite{Kim_2022}.

The thermal and magnetic properties of NCTO are anisotropic in nature and also the ground state magnetic spin structure is not yet clarified. Considering, these issues we have also performed the $I_{p}$  studies  on single crystal NCTO, under $H \vert \vert \,ab$ and $E \bot\,ab $  conditions, which also do not find any electric polarization. The earlier proposed zig-zag AFM spin structure of the NCTO is non-polar with and without magnetic field and the recently proposed triple-Q structure does not consist any specific point group. The triple-Q spin structure of NCTO would be expected to show the toroidal moment order parameter, which is odd, both under spatial inversion  and time reversal symmetries \cite{Spaldin,Gnewuch,VanAken,Gorbatsevich} and is  proportional to the antisymmetric part of the magnetoelectric tensor. Therefore, in such cases one would expect the linear increase of electric polarization as a function of the external magnetic field. However the present and earlier polarization studies do not show any such linear ME features in different directions of NCTO.

\section {Conclusion}

We have studied the in and out of plane susceptibilities of hexagonal NCTO single crystals and summarized the obtained C-W parameters along with the literature. Previously established in plane and out of plane susceptibility measurements mostly focus on the magnitude of the C-W temperatures, but the present study shows the obtained C-W temperatures qualitatively matches with theoretically calculated values. The temperature and magnetic field dependent dielectric studies on the single crystals at different temperatures below $T_{N}$ measured under $H \vert \vert \,ab$ and $E \bot\,ab $ directions shows a significant magneto dielectric coupling. The pyroelectric current studies on NCTO samples shows the absence of the spin driven electric polarization properties. The present magnetic, dielectric and ferroelectric studies  of the Na$_{2}$Co$_{2}$TeO$_{6}$ suggest the zig-zag AFM structure is more favorable than the triple-Q structure with AFM Kitaev interactions.

\hspace{1cm}

\section* {Acknowledgment}

This work is supported by the Science and Engineering Research Board-National Post-Doctoral Fellowship (PDF/2021/002536) funding. This study was also supported by the Ministry of Science and Technology, Taiwan, under Grants No.,NSTC-112-2112-M-110-018 and NSTC-113-2112-M—110-006. Author group from IIT Bombay acknowledges support of central measurement facilities (HR-XRD, SQUID MPMS) at their institution.


\bibliographystyle{apsrev4-1}
\bibliography{bib}
\end{document}